\documentclass[fleqn,12pt,twoside]{article}
\usepackage{espcrc1}

\usepackage{graphicx}
\usepackage[figuresright]{rotating}

\newcommand{\AmS}{{\protect\the\textfont2
  A\kern-.1667em\lower.5ex\hbox{M}\kern-.125emS}}

\title{Deeply Virtual Compton Scattering at Jefferson Lab, Results and Prospects}

\author{Latifa Elouadrhiri\\
Physics Division, Jefferson Lab, Newport News, Virginia, U.S.A}

\newcommand{\pEpg}{$\vec ep~\rightarrow~ep\gamma~$}
\newcommand{\Epg}{$ep~\rightarrow~ep\gamma$}

\newcommand{\pEpX}{$\vec ep~\rightarrow~epX~$}

\def\gev2{GeV$^2$}
\def\q2{$Q^2$}
\def\deg{$^{\circ}$}

\begin{document}

\maketitle

\begin{abstract}

Recent results from the Deeply Virtual Compton Scattering (DVCS) program
at Jefferson Lab will be presented. 
Approved dedicated DVCS experiments at 6 GeV and plans for the 12 GeV upgrade will be discussed.

\end{abstract}

\section{INTRODUCTION}

The recently developed formalism of ``Generalized Parton
Distributions'' (GPDs)  
\cite{Mul94,Ji97,Rady} 
showed that  information on quark-quark correlations, the transverse quark momentum distribution, and 
contributions of correlated quark-antiquark pairs (mesons) to the nucleon
wave function  can be obtained in hard \underline{exclusive}
 leptoproduction experiments.
 GPDs provide a unifying framework for the interpretation of
an entire set of fundamental quantities of hadronic structure, such as,
the vector and axial vector nucleon form factors, 
the polarized and unpolarized parton distributions, and the spin components 
of the nucleon due to orbital excitations. 
\smallskip
 
Deeply Virtual Compton Scattering (DVCS) is one of the key 
reactions to 
determine the GPDs experimentally, and it is 
the simplest process
 that can be described in terms of GPDs.
 The dominance of the handbag diagram and the behavior of the reduced
forward cross section as $1/Q^4$ (scaling regime) is expected to be
reached at lower
$Q^2$ than in the case of deeply exclusive meson production. 
This is supported by measurements of the
$\gamma^* \gamma \pi^0$ form-factor in $e^- e^+$ collisions \cite{CLEO}.

\smallskip

One of the  first experimental observation of DVCS was obtained from the recent  
analysis of CLAS data 
with a $4.2$ GeV polarized electron beam in a  kinematical regime 
near  \q2 = 1.5 \gev2 and $x_B~=~0.22$ \cite{SBE}. New measurements at higher 
energies are currently being analyzed, and dedicated experiments are planned. 
The high luminosity available for these measurements will
make it possible to determine
details of the $Q^2$, $x_B$, and $t$ dependences of GPDs.

\section{FIRST OBSERVATION OF EXCLUSIVE DVCS WITH THE CLAS DETECTOR}

\begin{figure}
\vspace{25mm} 
{\includegraphics{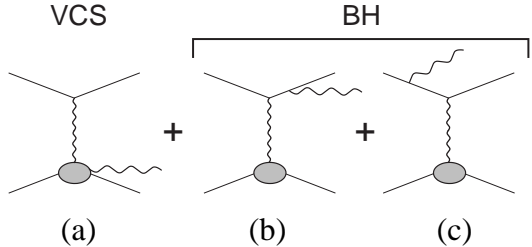}}
\caption
{Feynman diagrams for VCS and Bethe-Heitler processes
contributing to the amplitude of \Epg~ scattering.}
\label{bhdvcs}
\end{figure}

The DVCS/Bethe-Heitler (BH) interference has recently been measured using the CEBAF Large Acceptance Spectrometer in Hall B at Jefferson Lab \cite{SBE}. 
The data were collected as a by-product of the 1999 run with a 
4.25 GeV polarized electron beam. 
At energies above 4 GeV, the CLAS
acceptance covers a wide range of kinematics in the deep inelastic
scattering domain ($W \ge$ 2 GeV and $Q^2~\ge$ 1 \gev2). 
The open acceptance of CLAS and the use of a single electron trigger ensures 
event recording for all possible final states. This experiment measures DVCS via the interference
with the Bethe-Heitler   (figure \ref{bhdvcs}.
At beam energies accessible at Jefferson Lab, the BH contribution in
the cross section is predicted to be
several times larger than the DVCS contribution
in most regions  of the phase space.
The dominant BH process can be   turned into an advantage 
by using a longitudinally polarized electron beam:
one can measure the helicity-dependent
interference term that is  proportional to the 
imaginary part of the DVCS amplitude. In this case the pure real BH contribution 
is subtracted out in the cross section difference.

For the present DVCS analysis, electron and  proton were both detected in the CLAS detector, the reaction \pEpX was studied and
the number of single photon final states
was extracted 
by fitting the missing mass ($M_X^2$) distributions. 
The beam spin asymmetry was then calculated as:

\begin{figure}
\vspace{80mm} 
\centering
{\includegraphics{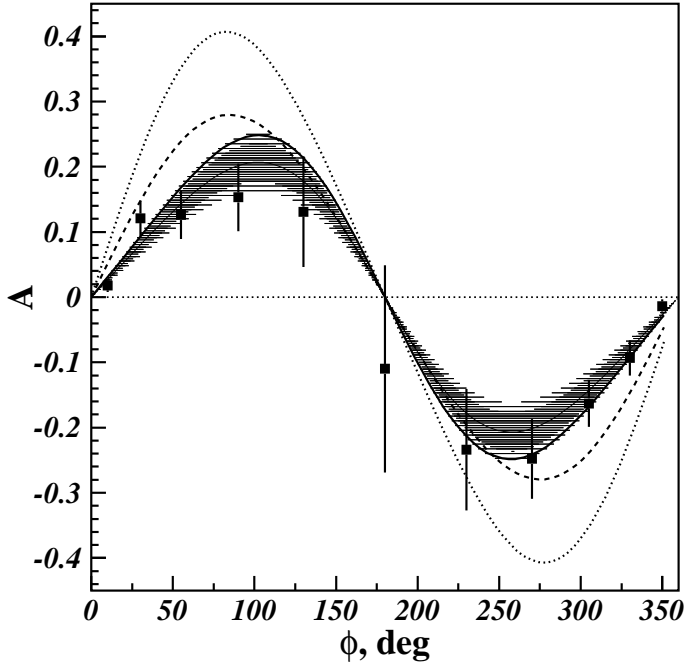}}
\caption
{\small $\phi$ dependence of the \pEpg beam spin asymmetry at 4.25 GeV. 
Data are integrated over the range of $Q^2$ from
1 to 2 \gev2, $x_B$ from 0.13 to 0.35 (with the condition $W>2$ GeV) 
and $-t$ from 0.1 to 0.3 \gev2. The shaded region is the range of the fit function A($\phi$)
defined by statistical and  systematical uncertanties.
The curves are model calculations according to Ref. \cite{MB}.}
\label{asym42}
\end{figure}

\begin{eqnarray}
{BSA~=~\frac{1}{P_e} \frac {\left(N^+_\gamma~-~N^-_\gamma\right)}
{\left(N^+_\gamma~+~N^-_\gamma\right)}}
\label{eq:bsan}
\end{eqnarray}
\noindent

\noindent
Here $P_e$ is the beam polarization and, 
$N^{+(-)}_\gamma$ is the extracted number of 
\pEpg  events at positive (negative) beam helicity.
The resulting $\phi$-dependence is shown in figure ~\ref{asym42}. A fit 
to the function

\begin{eqnarray}
{F\left( \phi \right)~=~A\sin\phi ~+~B\sin 2\phi }
\label{eq:asymf}
\end{eqnarray}

\noindent 
yields  $A~=~0.217 \pm 0.031$ and 
$B~=~0.027\pm 0.022$. If the handbag diagram dominates, as expected in the Bjorken regime,
$B$ should vanish 
and only the contribution from transverse photons should remain, described by
the parameter $A$. The GPD analysis including twist-3 contribution shows sensitivity of these data to 
$\bar{q}Gq$ correlations \cite{MB}.

\section{DEDICATED DVCS EXPERIMENTS AT JEFFERSON LAB}

There are two dedicated DVCS experiments planned to run using the 6 GeV polarized electron
beam. Both experiments plan to detect all three particles in the final state, the
scattered electron, the recoil proton, and the photon.

The first experiment E00-110~\cite{hallA} is a Hall A experiment which is expected to run in 2003. The DVCS beam spin
asymmetries and cross section differences will be measured at three 
$Q^2$ intervals, for a fixed interval of $x_B$. 
The experiment will provide a precise check of the 
$Q^2$ dependence of the
$ep\to ep\gamma$ cross section differences 
(for different beam helicities).

\noindent
The second experiment E01-113 \cite{HallB} is a dedicated CLAS DVCS experiment.  
The main goal of this experiment is to measure the $t$, $\phi$, and $x_B$ dependence
of the beam spin asymmetry for several fixed \q2 bins. 
This quantity is sensitive to the model description 
of the GPDs. This will be the 
first time this dependence will be  studied with high sensitivity using the DVCS process. 
A second goal will be to extract the helicity-dependent cross section 
difference, which directly determines the imaginary part of the
DVCS amplitude.

\begin{figure}
\vspace{70mm} 

{\includegraphics{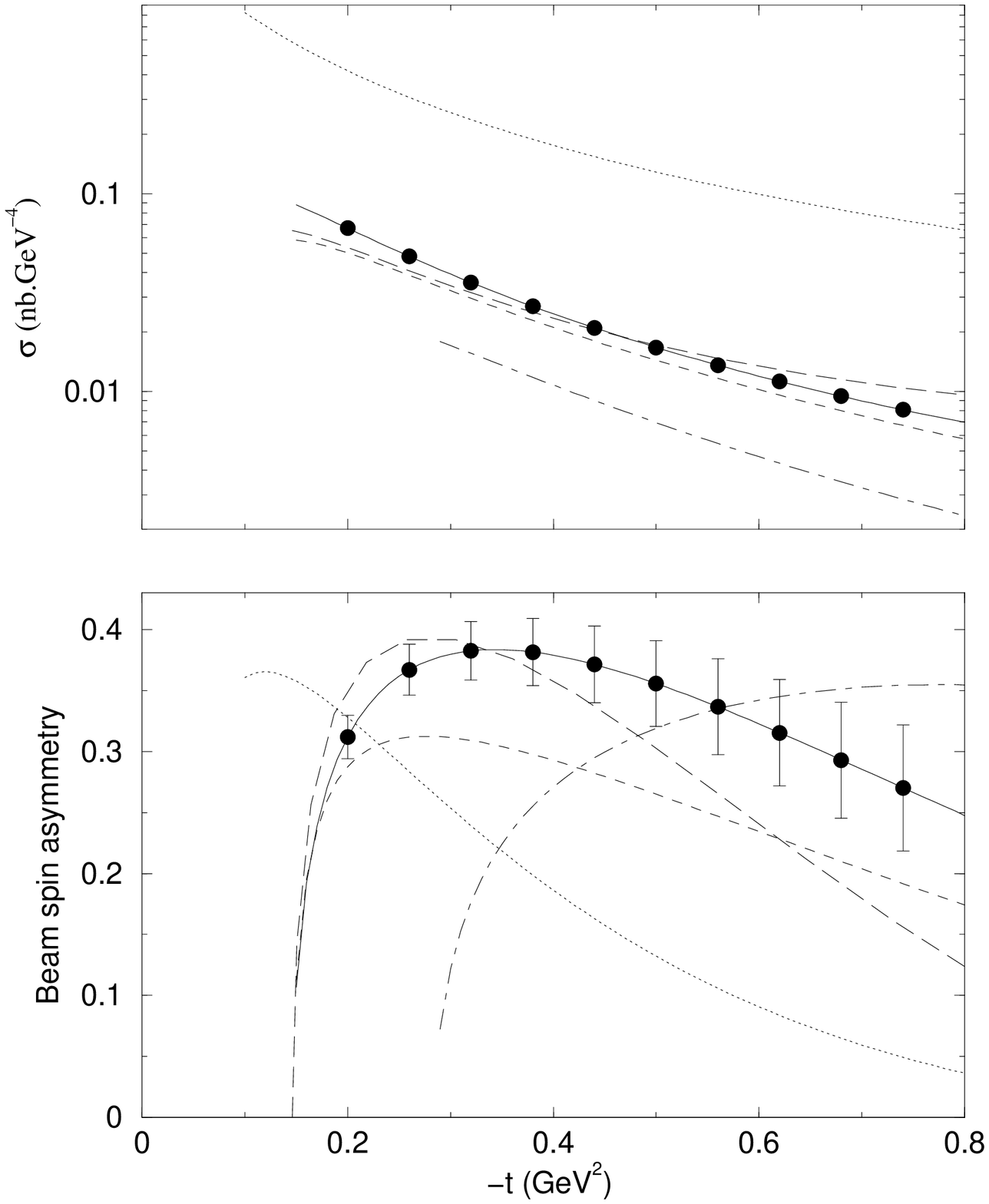}}
{\includegraphics{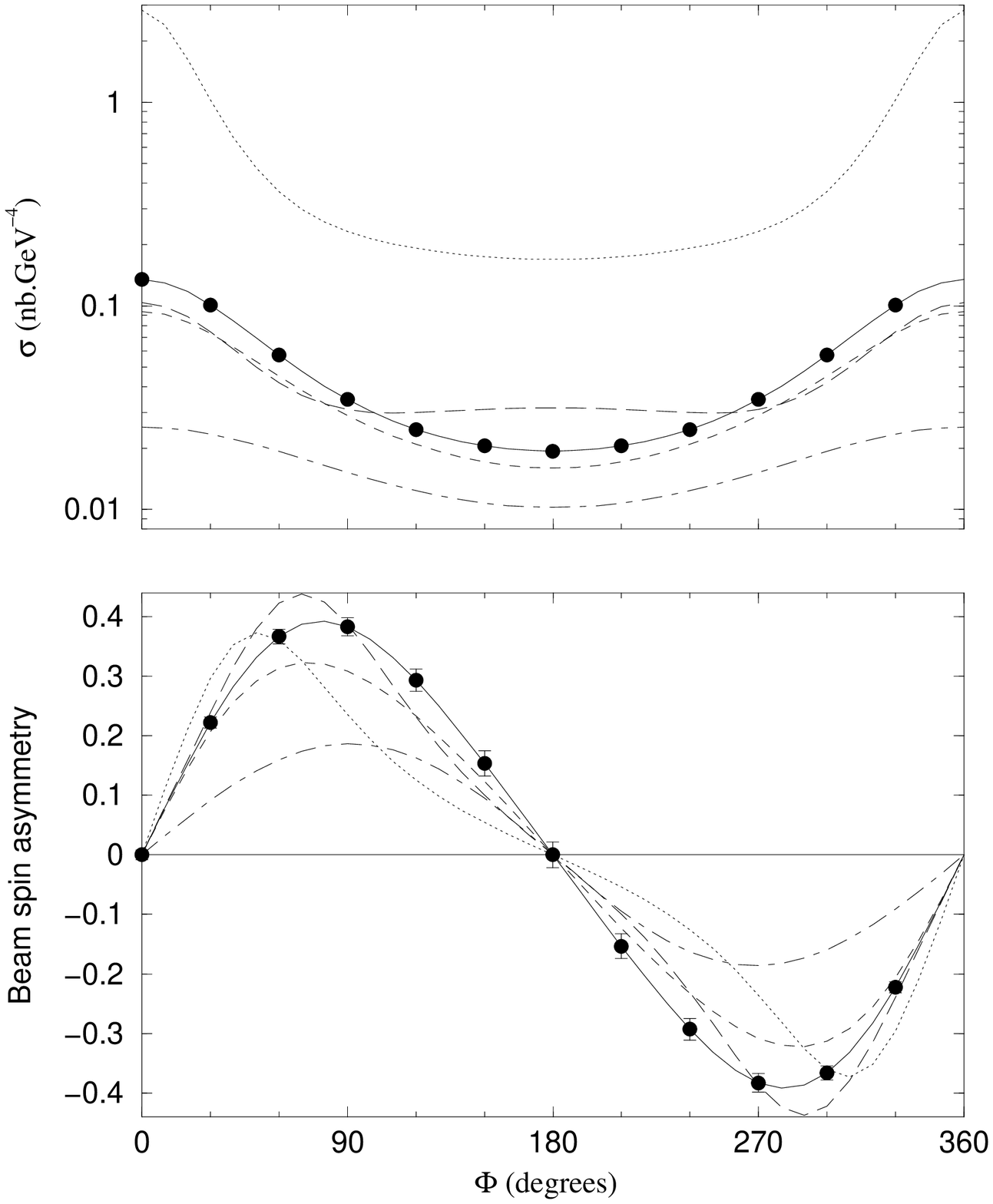}}
\caption{\small The left graph shows the $t$-dependence of \pEpg observables at 6 GeV, 
		for $\phi=90$\deg and
        \q2 = 1 \gev2, $x_B=0.22$ (dotted curve),
        \q2 = 2 \gev2, $x_B=0.35$ (solid),
        \q2 = 3.6 \gev2, $x_B=0.45$ (dot-dashed),
        calculated with the $\xi$-dependent GPDs of Refs.~\cite{GVG1,MG-COD}.
        From the same references, the $\xi$-independent version  is also shown
        (dashed),
        and from Ref.~\cite{Kiv00} the calculation including twist-3 effects 
        (long-dashed),
        both at \q2 = 2 \gev2.     
        The right graph shows  the $\phi$-dependence of \pEpg observables at 6 GeV, for  $t=-0.325$ \gev2.
        The points illustrate the expected statistical accuracy at \q2 = 2 \gev2. }
        
\label{tdep}
\end{figure}

Figure \ref{tdep} shows examples of the expected CLAS results, the $t$ and $\phi$ dependence of the cross section  
and of the beam spin asymmetry, for three bins  
in $Q^2$ and $x_B$. Expected data points are shown only
for $Q^2=2$ \gev2 and $x_B=0.35$.   
Comparison with different models for the $\xi$-dependence
of the GPDs is made, together with a first estimation
of the twist-3 effects for this process.

\smallskip

The results of these two experiments, Hall A and CLAS,  on the \pEpg cross section will allow
tests of the $Q^2$-dependence to check the scaling behavior.
As CLAS covers a broad kinematic range, we will be able to  
test the $Q^2$ dependence for different $x_B$.
This will verify that  we are in a regime where a direct 
interpretation of the results in terms of GPDs is possible. 
Observation of significant  scaling violations would provide 
important input for the analysis in terms of higher twist effects.

\section{CEBAF 12 GeV UPGRADE AND DVCS PROGRAM}

The study of GPDs via deeply exclusive reactions 
is one of the main research programs driving the CLAS upgrade to 12 GeV
\cite{WP}\cite{meck}.

\begin{figure}[ht]
\vspace{12.5cm} 
\includegraphics{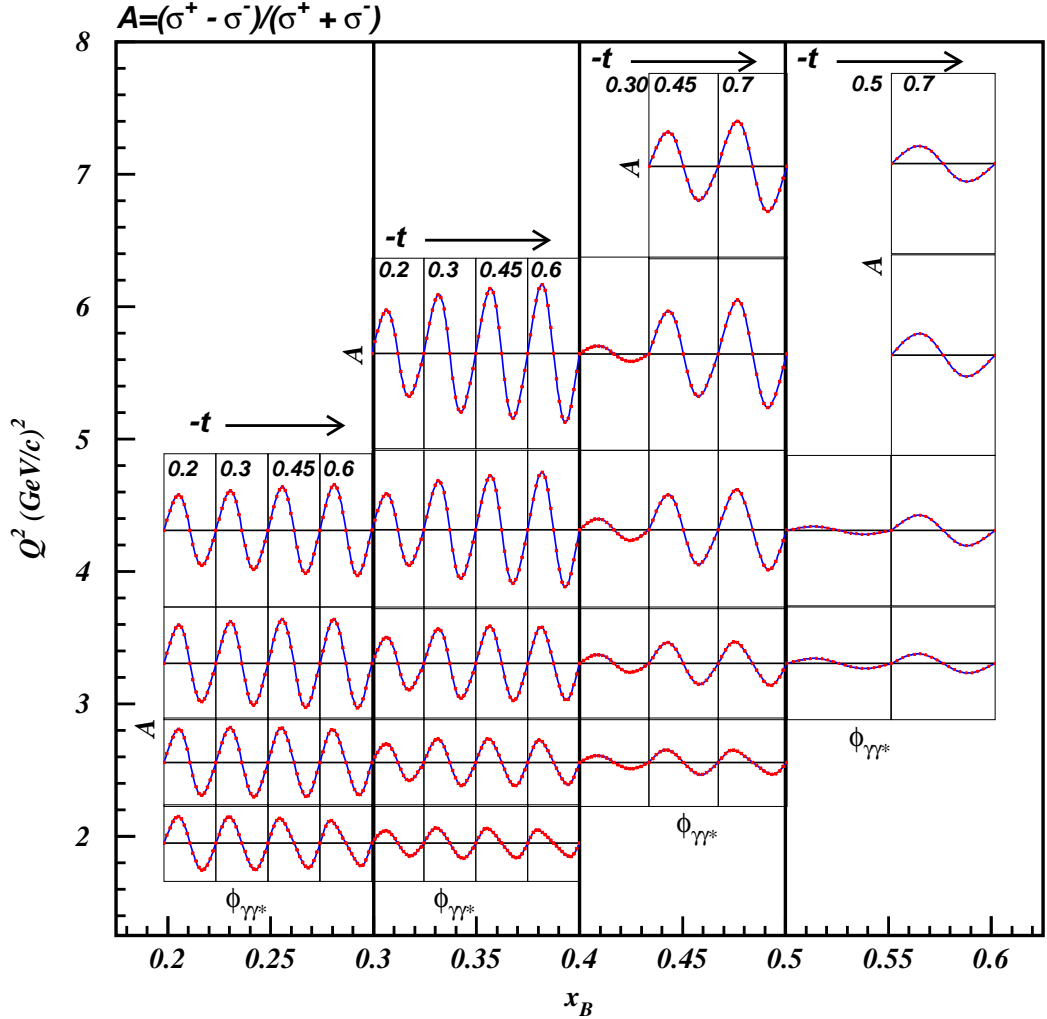}
\caption{\small{Kinematical coverage for beam spin asymmetry
measurements in DVCS. Possible binning of future data is shown.}}
\label{fig:all11}
\end{figure}

The DVCS cross section and spin asymmetries will be measured 
in a large range of  kinematical bins, simultaneously.
Figure \ref{fig:all11} illustrates the power of such measurements, showing
the expected statistical accuracy of the data, and an example of their binning. 
In the figure, simulated data on the beam spin 
asymmetry are shown in the 
56 bins of $x_B$, $Q^2$, and $t$ (a total of 1064 points).
The outer horizontal scale corresponds to  $x_B$,
divided into 4 bins, shown by solid lines. The outer vertical scale
represents $Q^2$. The $Q^2$ bins shown
as rows of asymmetry distributions. Larger bins at higher $Q^2$ are to
compensate for the fast
drop of the cross section. In each row, several
beam spin asymmetry distributions as a function of
$\phi_{\gamma\gamma *}$ (running from $0^\circ$ to $360^\circ$)
correspond to different $t$ bins. The average value of $-t$
is shown in the upper corner of each plot.

\begin{figure}[ht]
\vspace{9cm} 
\includegraphics{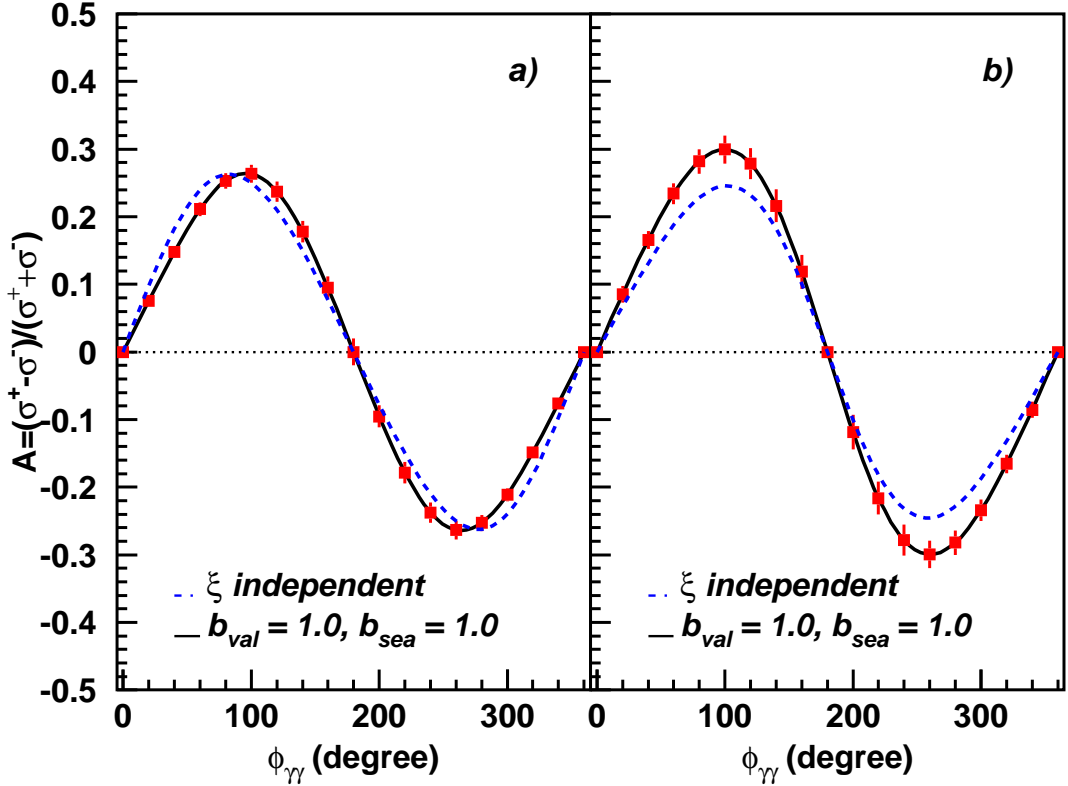}
\caption{\small{Expected data on the beam spin asymmetry at the
kinematics $x_B=0.35$ and $-t=0.3$ (GeV/c)$^2$, and a) 
$Q^2=2.75$ (GeV/c)$^2$ and b) $Q^2=5.4$ (GeV/c)$^2$. Expected
statistical errors are shown only. Data are simulated assuming 2000
hours of running at a
luminosity 10$^{35}$ cm$^{-2}$ sec$^{-1}$ with the upgraded CLAS
detector.}}
\label{fig:q2p2}
\end{figure}

Figure \ref{fig:q2p2} shows more clearly the accuracy of such a measurement and its
 sensitivity  to the models in 2 selected bins, $x_B=0.35$ and
$-t=0.3$ (GeV/c)$^2$, for $Q^2=2.75$ (GeV/c)$^2$ in
figure \ref{fig:q2p2}.a, and for $Q^2=5.4$ (GeV/c)$^2$
in figure \ref{fig:q2p2}.b.
As an input to the simulation
cross sections calculated in Ref.\cite{MG-COD} based on Ref. \cite{GVG1} have 
been used. 
Different curves in the figure correspond to different model
assumptions. One sees sufficient sensitivity for the separation of models
at both $Q^2$.  

\section{SUMMARY}

A first measurement of the beam
spin asymmetry in the 
exclusive electroproduction of real photons in the deep inelastic
regime was presented. 
We see a clear asymmetry, as expected from the interference of
the DVCS and BH processes.  It has been shown that our results can be accomodated 
within a  GPD analysis \cite{MB}. This supports the 
expectations that DVCS will allow access to GPDs at relatively low 
energies and momentum transfers.
This opens up a new
avenue for the study of nucleon structure which is inaccessible in
inclusive scattering experiments. 
Dedicated DVCS experiments at 6 GeV electron beam energy are planned, which will
allow significant
expansion of the $Q^2$ and $x_B$ range covered in these
studies. The high luminosity available for these measurements will
make it possible to map out
details of the $Q^2$, $x_B$, and $t$ dependences of GPDs.

The CEBAF 12 GeV upgrade will allow a breakthrough program, in the 
study of nucleon structure via deeply virtual exclusive reactions,
to be carried out with unprecedented precision.


\smallskip
\begin{thebibliography}{9}

\bibitem{Mul94}
D. M\"uller {\sl et al.}, Fortschr. Phys. {\bf 42} (1994) 2,101.


\bibitem{Ji97}

X. Ji, Phys. Rev. Lett. {\bf 78}, 610 (1997); Phys. Rev. D {\bf 55},
7114 (1997).

\bibitem{Rady}
A.V. Radyushkin, Phys. Lett. B {\bf 380}, 417 (1996); Phys. Rev. D
{\bf 56}, 5524 (1997).


\bibitem{CLEO} J. Gronberg et al. (CLEO Collaboration), Phys.Rev. {\bf D 57}, 33 (1998)  





\bibitem{SBE} S. Stepanyan et al., Phys.Rev.Lett., {\bf 87}
182002 (2001).

\bibitem{MB} A. Belitsky, D. Muller, and A. Kirchner, Nucl.Phys.,
{\bf B629}, 323 (2002).



\bibitem{hallA} P. Bertin, C. Hyde-Wright, F. Sabati\'e {\sl et al.},
		CEBAF experiment E00-110.


\bibitem{HallB} V. Burkert, L. Elouadrhiri, M. Gar\c con, S. Stepanyan {\sl et al.},
		CEBAF experiment E01-113.





\bibitem{GVG1} M. Vanderhaeghen, P.A.M. Guichon and M. Guidal,
Phys. Rev. {\bf D60}, 094017 (1999).


\bibitem{MG-COD} M. Vanderhaeghen, M. Guidal, P. Guichon, and L. Moss\'e, 
``Computer Codes for DVCS and BH
Calculations'', private communications.



\bibitem{Kiv00} N. Kivel, M.V. Polyakov, and M. Vanderhaeghen, Phys. Rev. D {\bf 63}, 114014 
(2001); and M.V., private communication.


\bibitem{WP} L.S. Cardman {\sl et al.}, The Science Driving the 12 GeV Upgrade of CEBAF, 2001


\bibitem{meck} B.A. Mecking, contribution to this conference.
 
\end{thebibliography}
\end{document}